\documentclass[aps, prl, reprint, showpacs]{revtex4-1}

\usepackage{graphicx}% Include figure files
\usepackage{dcolumn}% Align table columns on decimal point
\usepackage{bm}% bold math
\graphicspath{{./images/}}
\usepackage{amsmath}
\usepackage{epstopdf}

\usepackage[normalem]{ulem, xcolor}

\draft % marks overfull lines with a black rule on the right
\begin{document}

\title{Vertical-Cavity In-plane Heterostructures: Physics and Applications} %Title of paper

\author{Alireza Taghizadeh}
\author{Jesper M{\o}rk}
\author{Il-Sug Chung}
%\altaffiliation{}%Lines break automatically or can be forced with \\
\email{ilch@fotonik.dtu.dk} 

%\thanks{}
\affiliation{Department of Photonics Engineering (DTU Fotonik), Technical University of Denmark, DK-2800 Kgs. Lyngby, Denmark}

\date{\today}

\begin{abstract}
We show that the in-plane heterostructures realized in vertical cavities with high contrast grating (HCG) reflector enables exotic configurations of heterostructure and photonic wells. In photonic crystal heterostructures forming a photonic well, the property of a confined mode is determined by the well width and barrier height. We show that in vertical-cavity in-plane heterostructures, anisotropic dispersion curvatures plays a key role as well, leading to exotic effects such as a photonic well with conduction band like well and a valence band like barrier. We investigate three examples to discuss the rich potential of this heterostructure as a platform for various physics studies and propose a system of two laterally coupled cavities which shows the breaking of parity-time symmetry as an example. 
\end{abstract}

\pacs{42.55.Px, 42.55.Sa, 42.60.Da, 73.21.Fg}% insert suggested PACS numbers in braces on next line

\maketitle %\maketitle must follow title, authors, abstract and \pacs

\begin{figure}[tr]
\includegraphics[width=5.0cm]{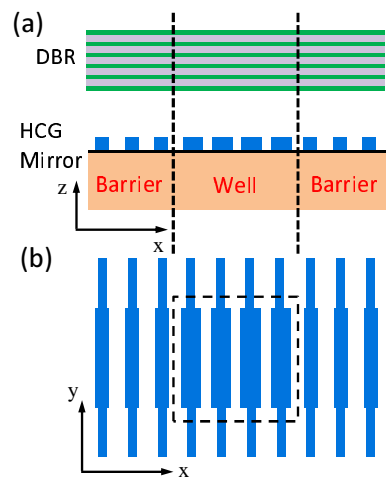}%
\caption{(a) Cross-sectional schematic of a vertical cavity structure with lateral variation of the cavity parameters in well and barrier regions. (b) Top view of HCG layer for a vertical-cavity in-plane heterostructures.
\label{fig:Schematic}}
\end{figure}

The vertical cavity is a rich platform for fundamental physics studies of light-matter interaction such as cavity quantum electrodynamics (QED) \cite{Gerard1998,Vahala2003} and cavity polaritons \cite{Deng2002,Christopoulos2007}, as well as various applications including vertical-cavity surface-emitting lasers (VCSELs) \cite{Moser2013, Westbergh2009}, single-photon light sources \cite{Pelton2002}, and Si-integrated on-chip lasers \cite{Chung2010}. Recently, it has been reported that the dispersion of a vertical cavity, i.e., the relation between the frequency $\omega$ and wavevector $k$ of a cavity mode, can be engineered by using the high-index-contrast grating (HCG) as reflector and designing the dispersion of HCG \cite{Wang2015, Zhang2015}. Furthermore, a heterostructure can be formed in $x$ direction by varying HCG parameters, as shown in Fig. \ref{fig:Schematic}(a) \cite{Sciancalepore2011}. In those reports, the discussions are focused on the dispersion profile in $k$ space along the $x$ direction or the profile of dispersion band edge in real space along the $x$ direction. The $x$ direction represents a direction perpendicular to the grating lines, as defined in Fig. \ref{fig:Schematic}. 

Here, we note the anisotropic dispersion curvatures of HCG-based vertical cavities along $x$ and $y$ directions and investigate their exotic impacts on the properties of vertical-cavity in-plane (VCI) heterostructures, which are not possible nor feasible in semiconductor and photonic crystal (PhC) heterostructures. The dispersion curvature is the second order derivative of the frequency of a propagating mode with respect to the in-plane wavevector and its inverse can be interpreted as an effective photon mass along the wavevector direction. 
As discussed below, two dispersion curvatures along $x$ and $y$ directions can be feasibly designed in HCG-based cavities to have a specific positive, zero, or negative value. Furthermore, a heterostructure with significantly different dispersion curvatures (even a different signs of curvatures) in each side of the heterostructure interface can be formed. 
%Thus, this enables control of frequency spacing of the different modes in a VCI heterostructure. 
In comparison, in semiconductors, the effective masses along different directions have values quite similar. In PhCs, since only the waveguide-type cavities are practically allowed for a laser cavity to avoid excessive surface recombination, only one effective mass is available along the light propagation direction. In addition, a heterostructure with opposite sign of effective masses in each side of the heterostructure interface is not possible for both semiconductor and PhC heterstructures. Therefore, with HCG-based VCI heterstructures, one may design a well structure with an extra degree of freedom of effective mass and a control over the direction of energy flow. These identified differences and consequent potentials have motivated the study of this paper.
 
This paper is organised as follows. Firstly, we derive an analytic expression for the dispersion of the entire cavity hereafter referred to as cavity dispersion. This expression decomposes the cavity dispersion into the contributions from each mirror and the waveguide that is a nominal cavity between two mirrors, hereafter referred to as mirror and waveguide dispersions, respectively. Using this expression, we quantitatively discuss the properties of anisotropic dispersion curvatures as well as an optimal thickness of the nominal cavity layer. Then, we investigate three characteristic VCI photonic well structures to study the physics of VCI heterostructures. In the first and second structures, all the well and barrier regions have positive and negative dispersion curvatures, respectively. These case studies explain how the order and spacing between the transverse mode frequencies are determined by the sign and magnitude of dispersion curvatures. As potential applications of these properties, the engineering of photon-photon resonances for achieving a high speed laser is discussed. In the third case, the well and barrier regions have opposite signs of dispersion curvatures, which shows the possibility of designing a heterostructure with another degree of freedom, compared to semiconductor or PhC heterostrucutres. Based on these new insights, we propose and analyze a novel system of laterally coupled vertical cavities. This appears promising for various physics studies where coupled laser cavities are needed, since the effective mass can be controlled as another degree of freedom and the coupling direction and light output/input direction are separated, which makes the access to the information in cavities much easier. As an example, the parity-time symmetry breaking phenomenon is reported for the first time in vertical cavities. 
%We believe that the unique possibility to HCG-based VCI heterostructure with various effective mass in well and barrier regions enables exotic photonic well configurations, which is not possible or feasible in semiconductor or PhC heterostructure. 

%Compared to a PhC heterostructure, in HCG-based VCI heterostructure, the dispersion can be changed abruptly at heterostructure interfaces without significant scattering. Furthermore, the status of a laser cavity can be easily readable or the laser cavity can receive an external input in a vertical direction, which is especially desirable for a network of coupled cavities. These properties provide more design flexibilities. Therefore, the HCG-based VCI heterostructure has a potential to be a rich platform for various physics studies which require accurate control of coupling between cavities, e.g., parity-time symmetry breaking phenomenon. These identified potentials have motivated the study of this paper.

\textit{Dispersions curvatures:} In \cite{Sciancalepore2011}, an expression for cavity dispersion was discussed around the Fano resonance frequencies of the HCG comprising the vertical cavity. Here, we derive a general expression which is valid at all frequencies within the HCG stopband, which is in line with discussion in \cite{Wang2015}. In a vertical cavity, the mode frequency $\omega$ is found by solving the oscillation condition:
\begin{equation}
\begin{split}
&\phi_1(\omega, k_x, k_y)+\phi_2(\omega, k_x, k_y)-2k_z L_c=2m\pi, \\
&k_z = \sqrt{\Big(\frac{n_c\omega}{c}\Big)^2-k_x^2-k_y^2}.
\end{split}
\label{eq:ResonanceCond}
\end{equation}
Here $k_x$, $k_y$, and $k_z$ are wavevector components of a mode in the nominal cavity layer with a refractive index $n_c$, which is a layer between mirrors, and $L_c$ is the nominal cavity length. Note that the reflection phases from the two mirrors, $\phi_1$ and $\phi_2$ depend on the in-plane wavevectors $k_x$ and $k_y$. The resonance frequency at normal incident denoted by $\omega_0$ is $2k_{z0} L_c$ = $-2m\pi+\phi_1(\omega_0,0,0)+ \phi_2(\omega_0,0,0)$, where $k_{z0}=\omega_0 n_c/c$.
Inserting the resonance condition into Eq. (\ref{eq:ResonanceCond}) and Taylor-expanding it lead to:
\begin{equation}
\label{eq:Dispersion}
\begin{split}
&\omega=\omega_0+\sum_{j=x,y}\beta_j k_j^2, \\ 
&\beta_j=\frac{c^2}{2n_c^2\omega_0}\frac{L_c}{L_\text{eff}}
             +\frac{c}{4n_c}\frac{1}{L_\text{eff}}\left(a_{1,j}+a_{2,j}\right),
\end{split}
\end{equation}
where $L_\text{eff}$ $(= L_c+L_1+L_2)$ is the effective cavity length, $L_i$ $(= -\frac{c}{2n_c} \partial \phi_i/\partial\omega)$ is the phase penetration into a $i$-th mirror, and $a_{i,j} = \partial^2 \phi_i/\partial k_j^2$. The $\beta_j$ represents the cavity dispersion curvature along $j$ direction. The first term of $\beta_j$ is the waveguide dispersion curvature for the round-trip propagation in the nominal cavity and always positive, while the second term of $\beta_j$ accounts for mirror dispersion curvature. The mirror dispersion curvature of HCGs can be either positive, negative, or even zero. In general, it is polarization sensitive and anisotropic. 
%Using Eq. (\ref{eq:Dispersion}), let us quantitatively  discuss the mirror dispersion curvatures of conventional distributed Bragg reflectors (DBRs) and HCGs, and their influences on the cavity dispersion curvatures. 
%Especially, we investigate the dispersion for two polarizations (TM an TE polarized light) and along two axial directions, $x$ and $y$.

%\begin{table}
%\caption{Estimated values of waveguide and mirror dispersion for two normal directions and two light polarizations. The values are estimated for a structure with an air cavity at $1550$ nm wavelength. For HCG cases, mirror structures with highly reflectivity (above $99.9 \%$) in more than $50$ nm bandwidth are considered.
%\label{tab:Dispersion}}
%\begin{ruledtabular}
%\begin{tabular}{ccccc}
%%Case & TM-x & TE-x & TM-y & TE-y\\
%Case & Waveguide & DBR & HCG\\
%  & $\frac{2L_c}{k_c}$ ($\mu m^{2}$) & $a_i$ ($\mu m^{2}$) & $a_i$ ($\mu m^{2}$)\\
%\hline
%Waveguide & $\simeq$0.2 & $\simeq$0.2 & $\simeq$0.2 & $\simeq$0.2 \\
%DBR & 0.01 to 0.05 & 0.01-0.05 & 0.01-0.05 & 0.01-0.05 \\
%HCG & 0.01-0.05 & 37.66345 & 86.37 \\
%TM-x & $\simeq$0.4 & 0.01 to 0.05 & -5 to 5 \\
%TE-x & $\simeq$0.4 & 0.01 to 0.05 & -10 to 10 \\
%TM-y & $\simeq$0.4 & 0.01 to 0.05 & -2 to 2 \\
%TE-y & $\simeq$0.4 & 0.01 to 0.05 & -1 to 1 \\
%\end{tabular}
%\end{ruledtabular}
%\end{table}  

%We have summarized the common range value of waveguide and common mirror dispersion in Table \ref{tab:Dispersion} and compare the properties of IPCD in DBR-based VCs with HCG-based ones in Table \ref{tab:Comparison}. 
At $\lambda_0$=1550 nm, a $0.5\lambda$-long nominal air cavity has a waveguide dispersion curvature of approximately 30 m$^2$/s. As seen in Eq. (\ref{eq:Dispersion}), it increases linearly with a longer cavity length $L_c$. The mirror dispersion curvature of a typical distributed Bragg reflector (DBR) is on the order of 1 m$^2$/s and gets smaller with a larger refractive index contrast of DBR layers. It is isotropic due to the rotational symmetry of DBR structures. Therefore, in DBR-based cavities the waveguide dispersion curvature dominates the cavity dispersion curvature and makes it always positive and isotropic, which resembles the conduction band in semiconductors. However, in the case of HCGs, the mirror dispersion curvature can be positive or negative and may be on the order of $\pm$100 m$^2$/s or even larger. If the cavity length $L_c$ is short, e.g., $1\lambda$ or 2$\lambda$, the cavity dispersion curvature of HCG-based cavities can be positive, zero or negative, being dominated by the HCG dispersion curvature. Also, the cavity dispersion is anisotropic along $x$ and $y$ directions and depends on the incident light polarization since the HCG mirror dispersion does. Therefore, it is possible in HCG-based cavities to engineer the cavity dispersion by designing the phase response of the HCG while keeping its reflectivity high. Designing of a phase is feasible for HCGs as discussed in \cite{Carletti2011}. We would note that a new type of grating reflector referred to as hybrid grating reflector \cite{Taghizadeh2014} or zero-contrast grating \cite{Magnusson2014} has similar mirror dispersion as HCG. The hybrid grating consists of a sub-wavelength grating layer and an unpatterned layer which may includes a gain material.% For this reflector, the in-plane mirror dispersion is similar to the HCG reflector. %It should be emphasized that for very long cavities, the waveguide dispersion always dominates.

The envelope approximation derived for PhC heterostructures \cite{Charbonneau2002, Istrate2006} can be applied to analyze VCI heterostructures \cite{Park2015}. The effective mass $m_j$ defined for the envelope approximation can be related to the dispersion curvature $\beta_j$: $1/m_j=\partial^2 \omega^2/\partial k_j^2=4\omega_0\beta_j$.
%It provides a graphical visualization of energy profiles of a heterostructure.
%When numerical simulations are needed, it can simplify a three dimensional (3D) problem with a smallest length scale of sub-micron to two dimensional problem with a smallest length of several microns. Thus, in designing vertical cavities, semi-quantitative simulation can be quickly done by using envelope approximation and fine designing can be done afterwards by using rigorous 3D simulations.
%By introducing a step-like change of the transverse properties at the boundary between the core and cladding section, a photonic well can be formed. 
%This change can be any geometrical change in the device dimensions or any carrier-induced change such as carrier-induced refractive index changes. In a conventional VCSEL with oxide aperture \citep{Bienstman2001}, the oxide aperture part works as the photonic well section. The cavity thus acts as a two-dimensional photonic well structure for in-plane direction. 
Using the envelope approximation, the resonance frequency of a rectangular VCI photonic well $\omega_{p,q}$ is found as \cite{Park2015}:
\begin{equation}
\omega_{p,q}^2 \simeq \omega_0^2+\frac{\alpha_x(p \pi)^2}{2m_xL_x^2}+\frac{\alpha_y(q \pi)^2}{2m_yL_y^2},
\label{eq:QWResonance}
\end{equation}
where $p$ and $q$ are mode numbers for the $x$ and $y$ directions, respectively, $\alpha_x$ and $\alpha_x$, rational factors due to the finite barrier heights, and $L_x$ and $L_y$, the lengths of the heterostructure in $x$ and $y$ directions, respectively. This expression will be used to interpret the result of example structures below.

\begin{figure}
\includegraphics[width=7.5cm]{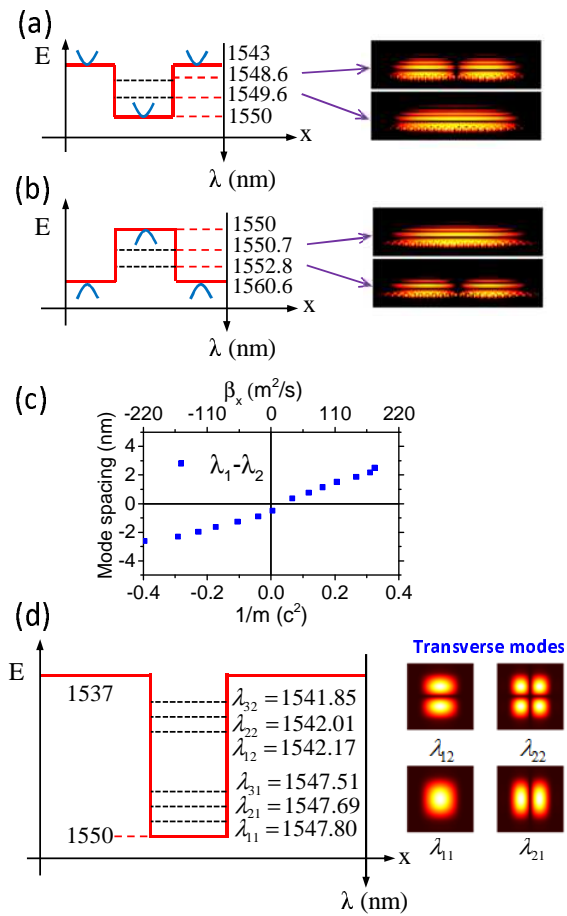}%
\caption{Transverse spatial variation of photonic bandedge for (a) positive and (b) negative in-plane dispersion, respectively. The first two transverse mode profiles are shown with their resonance wavelengths. Blue curves in (a) and (b) schematically represent dispersion curves. (c) Transverse-mode wavelength spacing between two lowest modes as a function of cavity dispersion band curvature. (d) Mode grouping effect in a two-dimensional photonic well due to different spatial extent or effective mass in x and y directions with corresponding 3D simulated transverse mode profiles of several lowest order modes.
\label{fig:ModeSpacing}}
\end{figure}

\begin{figure}
\includegraphics[width=5.5cm]{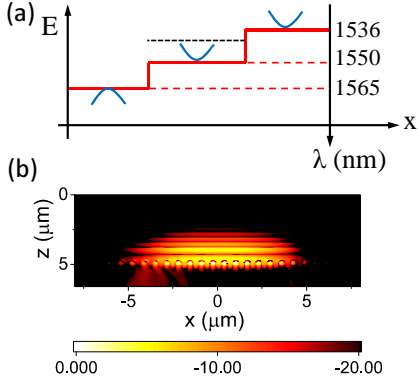}%
\caption{(a) Transverse spatial variation of photonic bandedge for a well structure with opposite effective mass in left and right barrier. (b) Normalized absolute value of magnetic field profile $H_y$ (logarithmic color scale) of the structure in Fig. (a). 
\label{fig:DifferentMass}}
\end{figure}

\textit{Three case studies:} To study the physics of VCI heterostructures, three characteristic photonic well structures are numerically investigated. As shown in Figs. \ref{fig:ModeSpacing}(a), \ref{fig:ModeSpacing}(b), and \ref{fig:DifferentMass}(a), the first structure has positive dispersion curvatures for both well and barrier regions, the second has negative dispersion curvatures for both well and barrier, and the third has a mixture of positive and negative dispersion curvatures. All structures have a 0.5$\lambda$-long air cavity and HCGs designed to be highly reflective for TM polarized field \cite{Chung2015}. For simulations, an in-house developed three-dimensional (3D)
simulator was used as explained in \cite{Taghizadeh2015}, which is based on rigorous coupled wave analysis (RCWA) method \cite{Moharam1995, Li1997} and employs absorbing boundary conditions \cite{Hugonin2005}. 

It is well known in the VCSEL literature that higher order transverse modes have higher frequencies, i.e., shorter wavelengths due to their higher in-plane wavevectors. However, the in-plane dispersion of HCG structures can significantly modify this characteristic property.  For the photonic well case with positive dispersion shown in Fig. \ref{fig:ModeSpacing}(a), we have the usual situation of VCSELs; the fundamental mode has the longest wavelength. However, for the negative dispersion case shown in Fig. \ref{fig:ModeSpacing}(b), the fundamental mode has a shorter wavelength than the higher order mode. Referring to Eq. (\ref{eq:QWResonance}), this observation can be interpreted like this: the higher order mode with more spatial modulation adds a larger negative kinetic energy, lowering the total energy. The positive dispersion (electron-like) and negative dispersion (hole-like) cases are analogous to the electronic quantum wells in conduction band and valence band, respectively. This observation is generalized in the result of Fig. \ref{fig:ModeSpacing}(c).

Figure \ref{fig:ModeSpacing}(c) plots the wavelength spacing of the two lowest transverse modes as a function of the $x$-direction dispersion curvature $\beta_x$ in the well region. It shows that the wavelength spacing increases with a larger dispersion curvature. This observation can be understood also by referring to Eq. (\ref{eq:QWResonance}): With a smaller dispersion curvature corresponding to a larger effective mass, the kinetic energy contribution to the total energy becomes smaller, leading to a smaller energy difference between two transverse modes. Therefore, the transverse mode spacing can be controlled by engineering the dispersion of HCG, without changing the transverse mode size. This results in several interesting phenomena such as mode grouping and mode degeneracy and has a potential to innovate several applications, as discussed below. 

If $m_x$ and $m_y$ largely differ, the transverse modes are grouped as shown in Fig. \ref{fig:ModeSpacing}(d). In Fig. \ref{fig:ModeSpacing}(d), the effective mass along $y$ direction is roughly 10 times smaller than that along $x$ direction. As a result, the second mode number for $y$ direction determines the larger wavelength spacing between groups, while the first one for $x$ direction determines the smaller wavelength spacing within a group. This mode grouping is experimentally observed in a HCG-DBR cavity laser \cite{Park2015} and 3D simulations give the same result. The mode profiles in Fig. \ref{fig:ModeSpacing}(d) are obtained by 3D simulations. Furthermore, the fundamental mode frequency $\omega_{0,0}$ and higher order mode frequencies $\omega_{p,p}$ can be made degenerate by designing the effective mass so that $m_x=-m_y$ and $\alpha_x/L_x^2=\alpha_y/L_y^2$ [c.f., Eq. (\ref{eq:QWResonance})]. This transverse mode-degeneracy is also confirmed by 3D simulations.

This control over transverse mode spacings can be used to boost the speed of a diode laser. Recently, the bandwidth boost of laser diodes has attracted lots of attention \cite{Graydon2015, Mieda2015}. The boost mechanism is based on the introduction of a photon-photon resonance at a frequency higher than the relaxation oscillation frequency by designing an external optical feedback and cross-gain modulation. In HCG-based vertical cavities, multiple transverse modes can be designed to have specific wavelength spacings, e.g., 0.15 nm, which determines the photon-photon resonance frequency. In this way, multiple photon-photon resonances can be introduced at designed frequencies through cross-gain modulation. For this, we need to make all transverse modes lasing, which is feasible since we can separately control the mode profile from gain profile, by using in-plane heterostructure. An external optical feedback in in-plane directions can be introduced as well. Recently, we have demonstrated a high-speed Si-integrated vertical cavity laser, using the in-plane optical feedback \cite{Park2015b}.

%In quantum wire cases with fixed electron mass, $L_x$ and $L_y$ need to be changed to implement mode grouping, while here we have the flexibility of changing the effective mass in the two in-plane directions. This phenomenon is verified by our rigorous 3D simulations and we have recently observed it experimentally \cite{Park2015}. 
%\new{[Comment: Didn't you comment this in the previous paragraph? You may decide whether to keep this sentence]}\delete{While in conventional vertical cavities employing DBR, different transverse modes have different resonance wavelengths due to their different wavevector components, this can be altered by the in-plane cavity dispersion. }

The unique possibility to design various effective masses in well and barrier regions enables exotic photonic well configurations. An interesting example is a photonic well where the effective mass sign of the barrier region is the opposite to that of the well region, e.g. hole-like barrier with electron-like well. In order to obtain a transverse mode confinement, the band edge of the barrier should be lower than that of the well. This band edge alignment is opposite to that of the Fig. \ref{fig:ModeSpacing}(a) case where both barrier and well are electron-like cavities. To compare these two cases, we design and simulate a VCI heterostructure with positive effective masses in the well and right barrier and negative effective mass for the left barrier, as  illustrated in Fig. \ref{fig:DifferentMass}(a). The right barrier is the same as in Fig. \ref{fig:DifferentMass}(a) case while the left barrier is the exotic configuration just described above. The mode profile is well confined in the well section, as shown in Fig. \ref{fig:DifferentMass}(b). If the left barrier band edge is moved above the well band edge, it results in unconfined mode. This example shows that HCG-based VCI heterostructure has more freedom to design photonic wells by controlling effective mass as well as band edge. %, we have more flexibility to control both %here it is possible to design the band edges and also effective masses to make heterostructure.\new{[Comment: Meaning is un-clear. You may change like this: This example shows a combination of effective mass and band edge frequency can be used for having mode confinement.]}

\begin{figure}
\includegraphics[width=8.5cm]{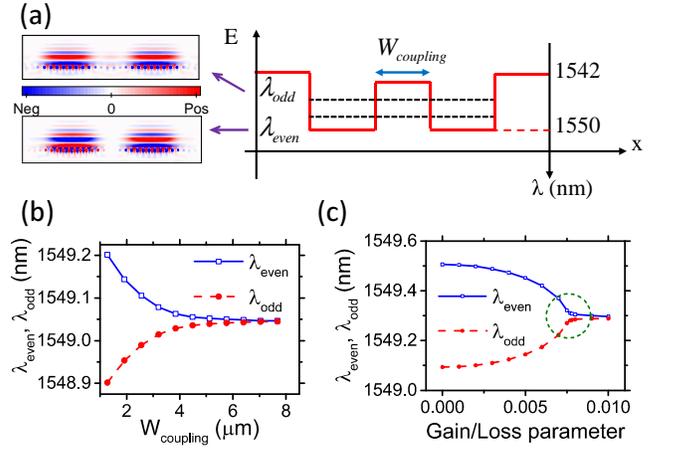}
\caption{(a) Transverse spatial variation of photonic bandedge for two laterally coupled vertical cavities with its odd and even modes. Real value of magnetic field profile $H_y$ of both modes are shown. (b) Resonant wavelengths of even and odd modes versus width of lateral spacing between two cavities. (c) Resonant wavelengths of of even and odd modes as function of gain/loss parameters which are the imaginary part of refractive indices. The green circle indicate the exceptional point where PT-symmetry is breaking.  
\label{fig:CoupledCavities}}
\end{figure}
%As we have shown in Eq. \ref{eq:DecayRate}, the penetration inside the barrier depend on its strength and the cavity dispersion.

Coupled micro-cavity structures are very attractive systems due to their physically rich characteristics resulting in many interesting phenomena such as miniband formation \cite{Happ2003}, heavy photons \cite{Bayindir2000}, coupled-cavity QED \cite{Hughes2007}, and recently parity-time (PT) symmetry breaking phenomena \cite{Peng2014}. For implementation, various structures have been suggested and investigated including PhC coupled cavities \cite{Happ2003}, microring resonators \cite{Peng2014, Jing2014} and vertically coupled VCSELs \cite{Stanley1994}. Here, we propose a new type of laterally coupled vertical cavities, as shown in Fig. \ref{fig:CoupledCavities}(a). 
We note that in the proposed coupled cavities the directions of the light propagation (vertical) and coupling (lateral) are separated from each other while they are in the same directions in other coupled cavities. This makes the access to the information of each cavity easier. As shown in Fig. \ref{fig:CoupledCavities}(a), the coupling of two identical cavities leads to two coupled states with even and odd parities. 
The coupling strength can be tuned by changing the barrier width, height, or effective mass. Here, we choose the barrier width, $W_\text{coupling}$. As shown in Fig. \ref{fig:CoupledCavities}(b), with a larger coupling, i.e., smaller $W_\text{coupling}$, the separation between two coupled state wavelengths becomes larger. As shown in Fig. \ref{fig:CoupledCavities}(c), the PT symmetry of this coupled cavities can be broken by introducing a gain region in one cavity and a loss region in the other cavity, which reduces the wavelength separation. The exceptional point as indicated by a green dotted circle is a characteristic signature of PT-symmetry breaking \cite{Peng2014}.

In conclusion, the in-plane dispersion curvatures of a HCG-based vertical cavities can be designed to be specific values in different directions provided that the cavity length is short, e.g., $1\lambda$ thick. We have shown that this provides another degree of freedom in designing vertical cavity in-plane heterostructures, enabling exotic heterostructures which are not possible nor feasible in semiconductor and photonic crystal counterparts, e.g. a  photonic well with opposite sign of effective mass for well and barrier. Furthermore, a system of two laterally coupled vertical cavities was proposed and simulated which shows spontaneously broken parity-time symmetry.

\begin{acknowledgments}
The authors gratefully acknowledge support from the Danish Council for Independent Research (Grant No. 0602-01885B), the Innovation Fund Denmark through the HOT project (Grant No. 5106-00013B), as well as Villum Fonden via the NATEC Centre of Excellence.
\end{acknowledgments}

%\bibliographystyle{apsrev4-1}
\bibliography{Hetrostructure}

\end{document}